# Price elasticity of electricity demand:
# Using instrumental variable regressions to address endogeneity and autocorrelation of high-frequency time series

**Silvana Tiedemann,**[a,*] **Raffaele Sgarlato,**[a] **Lion Hirth,**[a,b]

[a] Hertie School, Centre of Sustainability, Germany

[b] Neon Neue Energieökonomik GmbH, Germany

[*]corresponding author: tiedemann@hertie-school.org

Silvana Tiedemann and Raffaele Sgarlato contributed equally to this work.

**Abstract.** This paper examines empirical methods for estimating the response of aggregated electricity demand to high-frequency price signals, the short-term elasticity of electricity demand. We investigate how the endogeneity of prices and the autocorrelation of the time series, which are particularly pronounced at hourly granularity, affect and distort common estimators. After developing a controlled test environment with synthetic data that replicate key statistical properties of electricity demand, we show that not only the ordinary least square (OLS) estimator is inconsistent (due to simultaneity), but so is a regular instrumental variable (IV) regression (due to autocorrelation). Using wind as an instrument, as it is commonly done, may result in an estimate of the demand elasticity that is inflated by an order of magnitude. We visualize the reason for the Thams bias using causal graphs and show that its magnitude depends on the autocorrelation of both the instrument, and the dependent variable. We further incorporate and adapt two extensions of the IV estimation, conditional IV and nuisance IV, which have recently been proposed by Thams et al. (2022). We show that these extensions can identify the true short-term elasticity in a synthetic setting and are thus particularly promising for future empirical research in this field.

## 1. Introduction

**Electricity demand.** Electricity demand must always be in balance. Flexible generators adapt their production to meet the demand; demand, on the other hand, has long been assumed to be inflexible (Cramton et al., 2013). With the increasing share of renewable energy, the generation is increasingly determined by the weather, a non-controllable factor. The volatility of hourly electricity prices increases and thereby the incentive for demand to react to hourly prices. On a societal level, a more short-term elastic demand is also beneficial: it allows integrating higher shares of non-dispatchable renewable capacities, whilst also decreasing the need for back-up capacity, thus reducing cost (Bushnell, 2005; Fabra et al., 2021). In liberalized electricity markets, a flexible demand further reduces the ability of generators to exercise market power (Borenstein et al., 1999; Albadi and El-Saadany, 2008; Fabra et al., 2021)[1].

**Previous research.** Research that empirically estimates the short-term elasticity of electricity demand is sparse. Insights mainly result from experiments that may not be valid beyond the experimental setting, or studies that focus on the demand of one sector or individual consumers only (for a longer discussion see Hirth et al., 2023). While such studies are valuable for gaining first insights, they cannot generate a representative estimate for the aggregate demand elasticity as it is used as an input in various applications, including academic publications (e.g. Schittekatte et al., 2022), and commissioned studies about the need for market reforms or back-up capacities (e.g. ACER, 2020; ENTSO-E, 2022; NERC, 2022). Policy makers, independent system operators, and researchers alike, therefore, need a reliable quantification and monitoring strategy of the development of the elasticity of aggregated electricity demand at hourly granularity.

**Identification approaches.** Estimating elasticities is non-trivial because of an endogeneity problem: electricity prices and aggregated demand[2] are determined simultaneously. For over a century, instrumental variable (IV) regressions have been a standard approach to overcome endogeneity problems when estimating causal relationships. In the 1920th, the method was developed precisely for the estimation of elasticities (Angrist and Krueger, 2001), and has found extensive application in the electricity sector (Lijesen, 2007; Bönte et al., 2015; Eryilmaz et al., 2017; Knaut and Paulus, 2017; Fabra et al., 2021; Babii, 2022; Hirth et al., 2023). Instruments used by these authors include lagged prices, prices paid by consumers from other sectors, and weather variables which act either as supply or demand shifters.

---

[1] In Spain, for example, the regulator made real-time-pricing the default option for households to induce more flexible behaviour from the demand-side and thereby reduce market power by generators. The policy did not have the intended effect, as shown by Fabra et al. (2021).

[2] By aggregate demand we mean the entire demand in a price zone or at least a sector which has the potential to affect the price of electricity. The assumption of a "price taking demand", often made by empirical studies dealing with time series from individual consumers, is no longer justified.

**Autocorrelation.** The most common and convincing supply shifter is wind energy, as it fulfills key requirements of an instrument: it has a considerable influence on the electricity price (relevance) and is clearly independent of electricity demand (exogeneity) (Bönte et al., 2015; Knaut and Paulus, 2017; Fabra et al., 2021; Hirth et al., 2023). What has not been adequately discussed, however, is the extent to which its autoregressive properties could induce biases. In the context of overcoming omitted variable bias, the recent work by Thams et al. (2022) shows that dependencies in the past can lead to inconsistent IV estimates. We show in this paper that the same applies when estimating the elasticity of demand.

**Research questions.** This paper compares and evaluates IV identification strategies that are used for estimating the response of electricity demand to high-frequency price signals. Particularly, we investigate the statistical properties preventing regular IV regression from correctly identifying the demand response, and which alternatives can reduce or eliminate the identified inconsistencies. To this end, we create a controlled testing environment by generating synthetic time series for hourly electricity demand and prices. Each observation is an economic equilibrium where the price is determined by the intersection of the demand and the wind-driven supply curve. The autocorrelation coefficients of the demand curve are derived from hourly Danish industrial demand to yield realistic magnitudes.

**Results.** The use of external instruments, such as wind, is indeed necessary to accurately identify a price response. However, wind as an instrument can result in the overestimation of effect by an order of magnitude due to the autocorrelation. We argue that one may think of such estimates not as the instantaneous response of demand to prices (as it is usually presented), but instead as a cumulation of the price response over a longer period. To estimate the instantaneous response, Thams et al.'s extensions of the IV design, conditional IV and nuisance IV, allow for a precise identification of the causal effect. A valid specification, however, requires careful considerations about the autoregressive properties of wind and demand. We end by indicating promising avenues for further research.

## 2. Properties of electricity demand, and prices

### 2.1. General properties

**Price formation**. The hourly electricity price is the result of a market equilibrium between supply and demand, i.e., the price results from the intersection of the supply and the demand curve (Figure 1). The stability of the electricity grid requires that there is always a balance between supply and demand, and the possibilities to store electricity are limited. Hence, every shift in supply triggers an immediate alteration of the supply curve, and hourly electricity prices are thus highly volatile,[3] increasingly so because of the expansion of non-dispatchable renewables (e.g. Ciarreta et al., 2020; Maniatis and Milonas, 2022; Hosius et al., 2023).

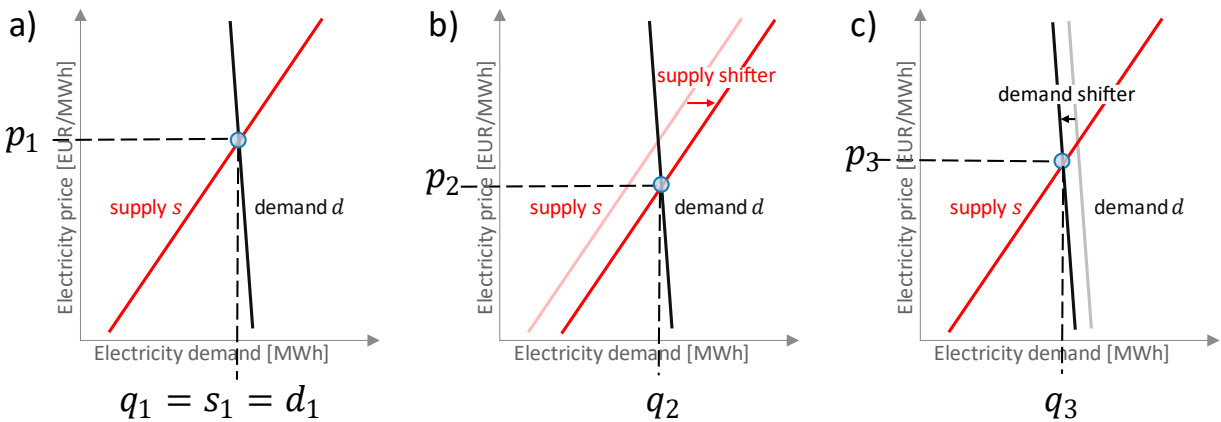

Figure 1: Electricity price formation. The hourly electricity price $p_t$ is determined by the intersection of the supply and demand curve, i.e., electricity prices and demand are determined simultaneously. In the electricity sector, the supply or merit order curve (red) features a stronger dependency on prices than the demand curve (black).[4] A typical supply shifter is wind or solar production. A typical demand shifter is demand for electric services that varies by hour-of-the-week.

**Supply**. The supply curve, in the electricity sector also referred to as merit order curve, ranks all available plants by their marginal cost of production. Typically, renewables are on the left with close to zero operational cost, or even negative "cost" if not producing is associated with foregone subsidies. Of these renewable sources, hydropower to some extent, but especially wind generation and solar generation are non-dispatchable. The rest of the supply curve is composed of mostly thermal generation sources, such as nuclear, coal, and gas fired plants. It is the production from thermal generators that compensate the stochastic nature of weather-driven renewable generation, especially in countries where hydropower is limited (Hirth, 2013).

---

[3] Note that volatility also depends on the size of the electricity market design (nodal vs. zonal) and the regional dimension of the price zone/interconnection. In price zones that cover a wider geographic area (e.g., France or Germany) the volatility should – a priori – be smaller than in countries with several price zones (e.g., Denmark and Sweden) or nodal systems (e.g., Texas).

[4] Strictly speaking, the figure represents the inverse of the demand and the supply curves. This representation is chosen because it is more common in economics.

**Demand.** The hourly demand for electricity exhibits a deterministic daily and weekly seasonality, even though there are interruptions to that pattern, for example for public holiday, during vacation periods, and – at least in predominantly Christian countries – Christmas time (Zweifel et al., 2017). Furthermore, the demand is altered by weather-dependent variables like temperature, precipitation, and sunlight. This leads to both inter-annual variation (e.g., cold vs. milder winters) and stochastic variation at very short time scales (e.g., a cloud covering the sun and increasing the demand for lighting). Furthermore, trends spanning multiple years exist. They are influenced by energy carrier substitution, as well as by energy efficiency and conservation measures.

**Autocorrelation.** Autocorrelation is ubiquitous in the electricity sector, even after controlling for the deterministic patterns and weather-dependent variables (see Figure 2). Processes requiring electricity often run for multiple hours. Examples can be found in all three major sectors, from industrial activities spanning an entire working shift, to commercial activities like water pumping or warehouse cooling, to private activities like washing machines. Hence, it is more likely that demand is high in one hour if the demand in the previous hour was high as well. The supply of electricity also depends on its immediate past. This is true for both conventional generators which are thermally inert and cannot adjust their production in incremental quantities, and for renewable electricity, whose production depends on the autocorrelated weather. However, for the sake of simplicity and because it should not affect the main arguments made in this paper, the autocorrelation of supply is disregarded in the following analysis.

## 2.2. Price elasticity

**Price exposure.** Historically, demand is assumed not to react to prices (Cramton et al., 2013) and there are several reasons why it makes sense to question a short-term demand elasticity. To be able to react, demand needs to be exposed to hourly prices. That is, the price signal that consumers face needs to figure at least some intra-daily variation. This is only the case if consumers are active on the electricity market directly (which tends to be the case of large consumers) or if consumers have retail contracts that are partially or fully indexed to wholesale prices, such as real-time-pricing or time-of-use tariffs that are correlated with wholesale prices. Furthermore, retailers would have an interest in influencing the demand of their customers also via non-price-based programs. The extent to which smart meters, flexible prices and incentive programs are available for small customers varies greatly in individual European countries, and is overall still limited (Tractebel, 2020).

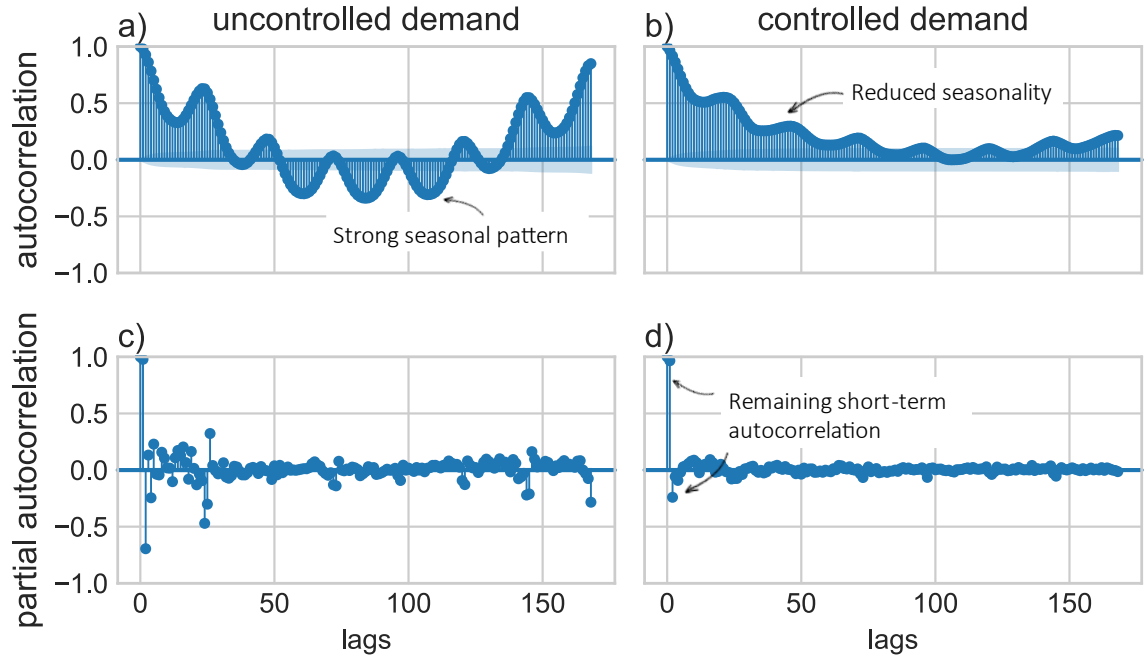

Figure 2: Autoregressive properties of electricity demand. Within a week (168 hourly lags), Danish industrial demand features strong seasonal autocorrelation visible in the pattern of the autocorrelation function (panel a) and the spikes of the partial autocorrelation function (panel c). After controlling, the periodicity reduces (panel b), but the autocorrelation of lags in the immediate past remains strong (panel d).

**Own-price and cross-price elasticity**. Once exposed to prices, consumers can react in different ways, only the first being the effect we are interested in (for a classification see Albadi and El-Saadany, 2008). First, they can react by changing the demand in the very same hour, which is commonly referred to as the own-price elasticity. Bönte et al. (2015) and Hirth et al. (2023) find a significant response of electricity traded on the main German power exchange and aggregated electricity demand in Germany, respectively, which Hirth et al. attribute to a response from the industrial sector. While Fabra et al. (2021) find that the individual residential consumers are not responding to prices. Second, consumers can react by shifting demand to other hours within the same day or across days (cross-price elasticity). For example, Taylor et al. (2005) and Zhou et al. (2019) find evidence that demand is shifted to hours that are distant in time, not to adjacent hours. As Zhou et al. (2019, p. 9) state „consumption in adjacent hours is complementary, while consumption in distant hours acts as a substitute". Third, the last way to react to prices is to substitute the electricity consumption with other energy carriers (substitution elasticity). While we recommend exploring the ability of IV designs to capture a cross-price elasticity in further extensions of this work, substitution is less likely to occur in the very short term.

**Bid curves**. A final note on the relationship between bid curves observed on electricity exchanges and the aggregated demand curve. Wholesale electricity is traded to a large extent in so-called day-ahead auctions with hourly granularity. On the day preceding the delivery, individual consumers or aggregators of small-scale consumption declare their willingness to pay for each hour. Simultaneously, suppliers submit their corresponding bid curves. The price corresponds to the intersection between the bids of suppliers and consumers. One could be tempted to compare empirical estimates of the elasticity of aggregate demand to observed bid curves. However, we argue that bid curves cannot be interpreted as the aggregated (not even the individual) demand curve, because part of the energy is traded on other markets and sold/bought with over-the-counter contract. Yet, prices remain representative because of the possibility of arbitrage between markets.

## 3. Identification strategies

### 3.1. (IV) strategies applied in electricity econometrics

**Objective.** Figure 3 represents the identification problem using a causal graph, the red arrow representing the causal effect $\beta^{(d)}$ of the price $p_t$ on demand $d_t$. A simple OLS regression ($d_t = \beta_0 + \beta^{(d)} p_t + \varepsilon_t$). would only find the compound effect between the elasticity and the back arrow from $d_t \rightarrow p_t$, thus suffering from the endogeneity bias caused by the simultaneity of prices and demand.

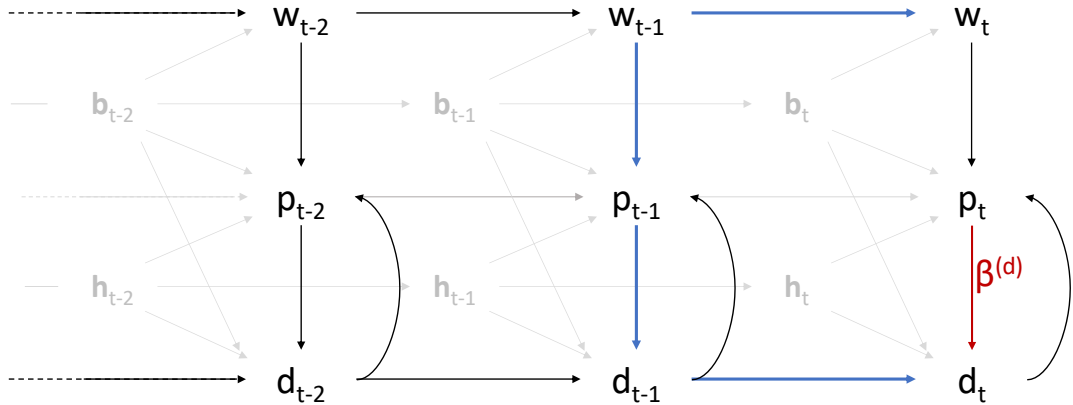

Figure 3: Causal graph. The causal effect of interest is $\beta^{(d)}$, representing the effect of the price $p_t$ on the demand $d_t$. If dependencies in time are disregarded, wind $w_t$ can be thought of as exogenous and therefore as a valid instrument. The autocorrelation of both demand and wind, however, open the blue path, which is shown to bias the estimate (since it violates the exclusion restriction).

**Instruments**. To overcome the simultaneity problem, economists have developed instrumental variable regressions (for a good overview see Angrist and Krueger, 2001). The idea is to replace the endogenous variable with an instrumental variable that is exogenous. Finding an instrument is difficult and usually requires an in-depth understanding of the existing dependencies (Cunningham, 2021). The instrument must be relevant, that is, it must have high explanatory power for the instrumented variable (Bound et al., 1995). Demonstrating the relevance is relatively easy. It is much more difficult to assess the validity of an instrument, which is, however, essential for obtaining unbiased results. To be valid, thus to fulfill the exclusion restriction, the instrument cannot affect the dependent variable through any other path than the variable being instrumented, and the assignment needs to be "as-good-as-random" – conditional on some control variables, if necessary (Angrist and Pischke, 2009).

**Wind.** In the electricity sector, the most common instruments are wind speed, the wind generation forecast, or wind generation (e.g. Bönte et al., 2015; Knaut and Paulus, 2017; Fabra et al., 2021;

Hirth et al., 2023). The authors argue that wind related instruments are valid because wind is a natural, non-controllable process, and fulfils the exclusion restriction (the most extensive discussion justifying wind as an instrument can be found in Hirth et al., 2023). As we will see, this does not hold true, if demand exhibits autocorrelation.

**Regular IV.** In terms of methodology, the two-stage least-squares regression (2SLS) is the most popular approach, where wind is used as the external instrument for the endogenous prices in the first stage (equation 1). The price estimate is then used in the 2nd stage to obtain the estimate for the slope of the demand curve $\beta^{(d)}$ (equation 2). We call this identification strategy *regular IV*.[5,6]

Regular IV: $\qquad$ 1st stage: $p_t = \beta_0^{(1)} + \beta^{(w)} w_t + \varepsilon_t$ $\qquad$ (1)

$\qquad\qquad\qquad$ 2nd stage: $d_t = \beta_0^{(2)} + \beta^{(d)} \hat{p}_t + \varepsilon_t$ $\qquad$ (2)

**Price as instrument.** An alternative approach is to use the lagged endogenous variable as an instrument, that is $p_{t-1}$. We label this identification strategy as *lagprice IV*, see equation (3) and (4). The argument of advocates is that the lagged price correlates sufficiently with the current price, yet the current demand cannot influence past realizations of the price. Such a design is "problematic if the equation error term or omitted variables are serially correlated" (Angrist and Krueger 2001, p.77). We still include this formulation in the comparison because it is used in the electricity sector, for example in the often cited work by Lijesen (2007).

Lagprice IV: $\qquad$ 1st stage: $p_t = \beta_0^{(1)} + \beta^{(p)} p_{t-1} + \varepsilon_t$ $\qquad$ (3)

$\qquad\qquad\qquad$ 2nd stage: $d_t = \beta_0^{(2)} + \beta^{(d)} \hat{p}_t + \varepsilon_t$ $\qquad$ (4)

### 3.2. Bias from autocorrelation

**Autocorrelation.** Thams et al. (2022) have shown that instruments used to overcome omitted variable bias can still lead to biased results if the instruments are autocorrelated (even if apparently exogenous). The Thams bias for an AR(1) process is proofed to be $\hat{\beta} = \beta / (1 - \alpha^{(i)} \alpha^{(d)})$, where $\alpha^{(i)}$ and $\alpha^{(d)}$ are the autocorrelation coefficients of the instrument and the dependent variable respectively. It is reasonable to suspect that a similar problem exists when endogeneity occurs through the simultaneous determination of prices and demand: Figure 3 shows that the autocorrelation of the instrument opens the path $w_t \leftarrow w_{t-1} \rightarrow p_{t-1} \rightarrow d_{t-1} \rightarrow d_t$ on which $w_{t-1}$ acts as a confounder, leading to a violation of the exclusion restriction.[7] We would therefore expect a biased result from a regular IV design, exception made for the cases discussed in the following paragraph.

---

[5] As a point of comparison, we include a version of the regular IV design where we apply a simple first difference operator, which should remove part of the autoregressive properties (e.g. Hirth et al., 2023).

[6] All identification strategies involving instruments are implemented using the Python package linearmodels, in particular the IV2SLS module (contained in linearmodels.iv.model; https://pypi.org/project/linearmodels/). The error estimation is always based on the Bartlett kernel for heteroskedasticity and autocorrelation robust inference (Andrews, 1991).

**Exceptions.** A regular IV design should lead to consistent estimates of the demand response in three special cases (Figure 4): (a) If demand is inelastic there is no link between $p_t \rightarrow d_t$. No open paths remain, and the estimate cannot be biased. Furthermore, if either (b) wind is purely random, that is the autocorrelation of the instrument is removed, or (c) demand does not feature autoregressive properties, the links between $w_t \leftarrow w_{t-1}$ and $d_t \rightarrow d_{t-1}$ vanish and the source of inconsistency is eliminated.[8]

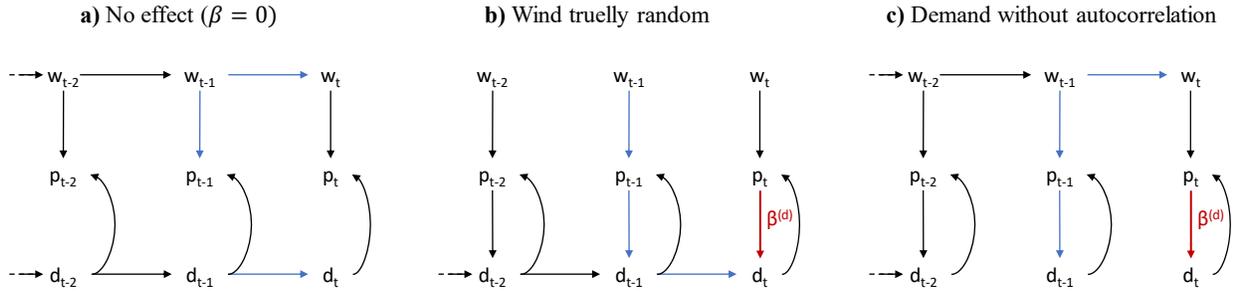

Figure 4: Causal graphs leading to unbiased regular IV designs. a) Demand is inelastic ($\beta^{(d)} = 0$), therefore the link $p_t \rightarrow d_t$ is missing. The path $w_t \leftarrow w_{t-1} \rightarrow p_{t-1} \leftarrow d_{t-1} \rightarrow d_t$ is closed because $p_{t-1}$ acts as a collider. b) Wind is truly random, i.e., the autocorrelation is removed. c) Demand is not autocorrelated. In cases b) and c) the blue path vanishes, and the exclusion restriction holds because the instrument $w_t$ only influences the dependent variable $d_t$ via the instrumented variable $p_t$. Note that this is true even if a path between $p_{t-1}$ and $p_t$ were to exist.

---

[7] More precisely, wind is not "as-good-as-random" anymore.

[8] It is worth noting that a randomly assigned instrument should also be valid if we were to assume an additional source of autocorrelation of supply: Wind $w_t$ at time t would still affect demand only through the price $p_t$.

## 3.3. Alternative specifications

**Blocking the path.** Thams et al. (2022) develop two identification strategies which block selected paths by conditioning on lagged variables. To the best of our knowledge, neither have found application in electricity econometrics yet.

- *Conditional IV* controls for the lagged instrument (wind), where M is the autoregressive order of wind:

$$1^{st} \text{ stage: } p_t = \beta_0^{(1)} + \beta^{(w)} w_t + \sum_{i=1}^{M} \alpha_i^{(w)} w_{t-i} + \varepsilon_t \quad (5)$$

$$2^{nd} \text{ stage: } d_t = \beta_0^{(2)} + \beta^{(d)} \hat{p}_t + \sum_{i=1}^{M} \alpha_i^{(w)} w_{t-i} + \varepsilon_t \quad (6)$$

- *Nuisance IV* controls for the lagged dependent variable (demand), where L is the autoregressive order of demand:

$$1^{st} \text{ stage: } p_t = \beta_0^{(1)} + \beta^{(w)} w_t + \sum_{l=1}^{L} \alpha_l^{(d)} d_{t-l} + \varepsilon_t \quad (7)$$

$$2^{nd} \text{ stage: } d_t = \beta_0^{(2)} + \beta^{(d)} \hat{p}_t + \sum_{l=1}^{L} \alpha_l^{(d)} d_{t-l} + \varepsilon_t \quad (8)$$

**Nuisance and conditional IV.** Both strategies require that one adjusts the number of control variables in line with the autoregressive degree of the process, and there is a qualitative difference resulting from the observable information. Wind is external to the system. It is therefore possible to observe the autoregressive degree and determine the "right" number of lags which are to be included, $M$. The autoregressive properties of the demand are more difficult to estimate. If one were to wrongly assume an insufficient autoregressive order $L$, paths could remain open. For example, using conditional IV with one lag of wind only, i.e., conditioning on $w_{t-1}$, only closes the blue path in in Figure 5 b), yet the orange one remains open because the instrument is an AR(2) process.

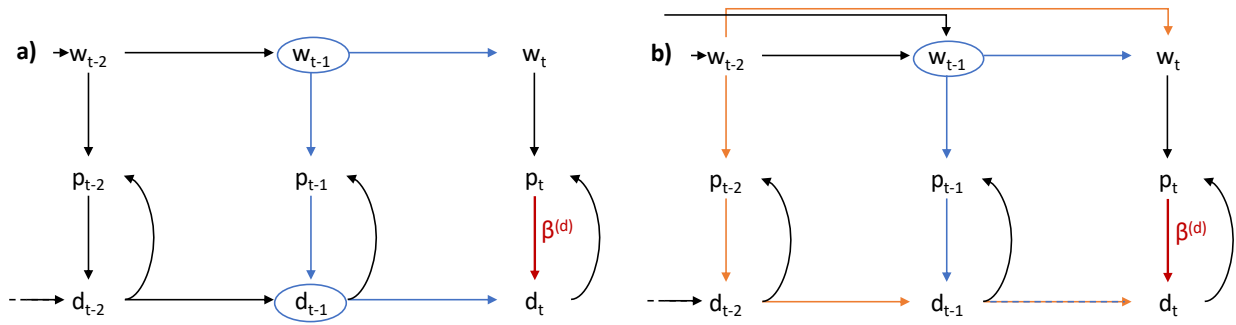

Figure 5: Conditional and nuisance IV. a) The idea of conditional and nuisance IV is to block the blue path inducing bias from autocorrelation by conditioning on either $w_{t-1}$ (conditional IV) or $d_{t-1}$ (nuisance IV). b) Conditioning for $w_{t-1}$ only (conditional IV) cannot block the orange path, because wind is an AR(2) process. Conditioning on $d_{t-1}$ (nuisance IV) can, because demand is an ARX(1) process.

## 4. Synthetic data generation

The main idea of the paper is to establish a synthetic data generation process that simulates the main features encountered in the electricity market and analyze to what extent IV identification strategies can correctly estimate the ex-ante defined price elasticity. The main features to encode are the simultaneous determination of price and demand, and the autoregressive properties of demand. We do not include additional features like seasonal patterns and trends, because an empirical strategy could easily control for these features.

**Demand equation**. Demand is modeled as an autoregressive process (equation 9). The goal is to mimic observed demand as close as possible. Therefore, we use the hourly electricity demand of the Danish manufacturing sector between the 1st of January 2019 to and the 28th of January 2020 and control for the daily, weekly, and monthly seasonality, for holidays, temperature, sunlight hours, and gas prices (Figure 2). We model the residuals with an AR model by restricting the autoregressive order to zero, one, or two. The coefficients that we obtain from this procedure are summarized in Table 1.

**Linearity.** The dependency of the price is assumed to be linear. We make this assumption for the sake of simplicity and because results from previous studies suggest that at least the functional form has little impact on the estimate (Hirth et al., 2023; Li et al., 2022)[9]. The slope of the demand curve $\beta^{(d)}$ is the own-price elasticity of demand. We alter its value to either represent a price-insensitive demand curve of slope $\beta^{(d)} = 0$ or a price-sensitive demand with a slope of $\beta^{(d)} = -0.4$. Furthermore, we add a constant, such that the average synthetic demand reflects the Danish industrial demand, and an error term $\varepsilon_t^{(d)}$, which is assumed to follow an independent, and identically distributed normal distribution with zero mean.

$$d_t := \beta_0^{(d)} + \beta^{(d)} p_t + \sum_{l=1}^{L} \alpha_l d_{t-l} + \varepsilon_t^{(d)} \qquad (9)$$

---

[9] Some empirical studies suggested that there is a heterogeneous response between off-peak and peak hours and between seasons (Frondel et al., 2019; Hirth et al., 2023), and that consumers may not respond until prices reach a critical threshold (Archibald et al., 1982; Damien et al., 2019; Shen et al., 2021; Li et al., 2022; Sloot and Scheibehenne, 2022). To identify the heterogeneous response, the studies often split the time series and estimate the response independently and/or include interaction effects. We leave the implementation of such a response for further research.

| Demand | L | $\beta_0^{(d)}$ | $\alpha_1$ | $\alpha_2$ | $Var(\varepsilon_t^{(d)})$ |
|---|---|---|---|---|---|
| AR(0) | 0 | 399.44 | - | - | 309.49 |
| AR(1) | 1 | 38.09 | 0.97 | - | 20.72 |
| AR(2) | 2 | 41.15 | 1.20 | -0.24 | 19.52 |

Table 1: Coefficients of the demand equation for different orders of autoregressions. The lags used to build the synthetic series affect the variance of the error, since larger lagged coefficients leads to smaller individual errors, that on the other hand, cumulate more.

**Supply equation**. The supply $s_t$ is also assumed to be a linear function of the price $p_t$. Furthermore, supply depends on the wind speed $w_t$. This corresponds to the empirical observation that (more) wind generation shifts the supply curve (outward).

$$s_t := \beta_0^{(s)} + \beta^{(s)} p_t + \beta^{(w)} w_t + \varepsilon_t^{(s)} \tag{10}$$

The intercept of the supply curve is set to zero, the effect of price variations is set to 4.0 MWh/EUR ($\beta^{(s)} = 4$), whereas the effect of wind speed variations is set to 16 MWh/ms$^{-1}$ ($\beta^{(w)} = 16$). We use empirically observed wind speeds in Denmark from 01-01-2019 to 28-01-2020[10], which is on average 7.4 and ranges from 2.2 to 18.3 ms$^{-1}$. The non-linear relationship between wind speed and wind generation is not considered. A small noise is added, $\varepsilon_t^{(s)} \overset{iid}{\sim} N(0, 0.01)$. The noise does not impact results but preserves the computational stability of the estimators.

**Equilibrium**. The price is determined by the market equilibrium where demand and supply meet. This point can be analytically described when the condition $d_t = s_t$ is expanded and rearranged. For readability, the operator $\Delta$ is used to describe the difference between a demand and the supply parameter $x$ ($\Delta x := x^{(s)} - x^{(d)}$). Figure 6 exhibits the resulting price equilibria. Table 2 gives summary statistics for the synthetic time series.

$$p_t = \frac{\Delta \beta_0 + \beta^{(w)} w_t - \sum_{l=1}^{L} \alpha_l d_{t-l} + \Delta \varepsilon_t}{-\Delta \beta} \tag{11}$$

---

[10] While we use 13 months for the date generation, we only use the data of one year for the estimation, dropping the observations of the first 28 days during which our data "swings in".

| Time series | | # | Mean (Std) | Min | Max | AR Lag 1 | Lag 2 |
|---|---|---|---|---|---|---|---|
| | Wind in m/s | 9432 | 7.6 (2.4) | 2.2 | 18.3 | 1.84 | -0.85 |
| Equilibrium | Demand in MWh | 8760 | 374.2 (17.4) | 310.4 | 434.1 | 0.02 | 0.04 |
| (L = 0, β = 0) | Price in EUR/MWh | 8760 | 63.7 (10.1) | 22.1 | 92.2 | 0.45 | 0.41 |
| | Wind in m/s | 9432 | 7.6 (2.4) | 2.2 | 18.3 | 1.84 | -0.85 |
| Equilibrium | Demand in MWh | 8760 | 373.3 (17.7) | 315.8 | 451.0 | 0.98 | -0.01 |
| (L = 1, β = 0) | Price in EUR/MWh | 8760 | 63.4 (10.2) | 26.5 | 92.1 | 1.32 | -0.34 |
| | Wind in m/s | 9432 | 7.6 (2.4) | 2.2 | 18.3 | 1.84 | -0.85 |
| Equilibrium | Demand in MWh | 8760 | 374.3 (17.7) | 318.1 | 430.5 | 1.19 | -0.24 |
| (L = 2, β = 0) | Price in EUR/MWh | 8760 | 63.7 (10.0) | 25.2 | 94.6 | 1.44 | -0.46 |
| | Wind in m/s | 9432 | 7.6 (2.4) | 2.2 | 18.3 | 1.84 | -0.85 |
| Equilibrium | Demand in MWh | 8760 | 373.9 (16.0) | 306.8 | 440.0 | 0.08 | 0.05 |
| (L = 0, β = −0.4) | Price in EUR/MWh | 8760 | 63.6 (9.1) | 25.6 | 92.7 | 0.48 | 0.39 |
| | Wind in m/s | 9432 | 7.6 (2.4) | 2.2 | 18.3 | 1.84 | -0.85 |
| Equilibrium | Demand in MWh | 8760 | 373.4 (26.2) | 308.2 | 462.4 | 1.04 | -0.06 |
| (L = 1, β = −0.4) | Price in EUR/MWh | 8760 | 63.4 (4.7) | 44.6 | 78.9 | 1.21 | -0.26 |
| | Wind in m/s | 9432 | 7.6 (2.4) | 2.2 | 18.3 | 1.84 | -0.85 |
| Equilibrium | Demand in MWh | 8760 | 373.1 (25.7) | 307.7 | 467.3 | 1.25 | -0.27 |
| (L = 2, β = −0.4) | Price in EUR/MWh | 8760 | 63.4 (4.6) | 40.6 | 77.7 | 1.33 | -0.39 |

Table 2: Summary statistics of the observed time series after the equilibrium condition was applied.

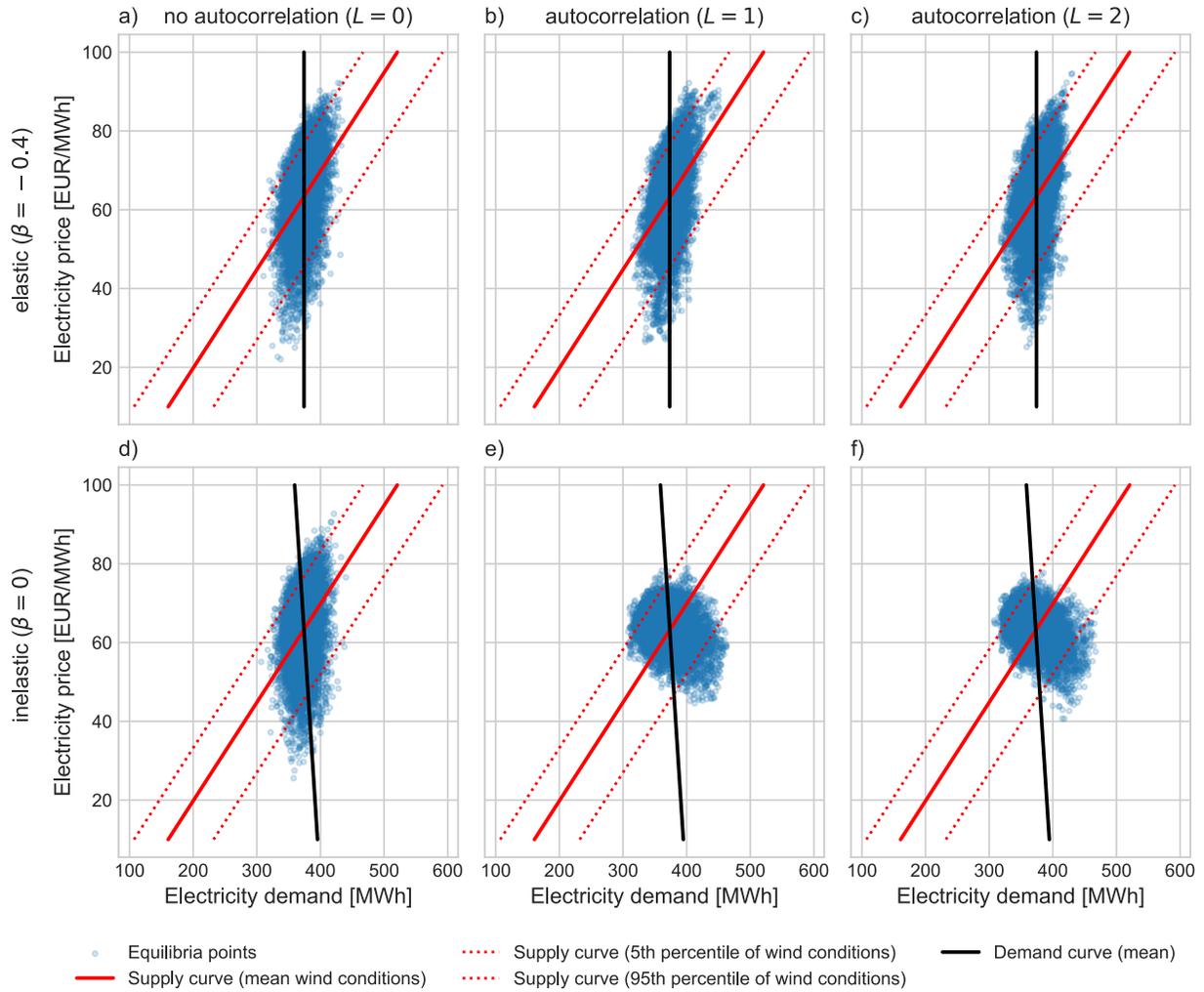

Figure 6. Correlation between price and demand for six synthetic data generations. The first row (panels a-c) shows equilibria resulting from inelastic demand ($\beta^{(d)} = 0$) In the second row (panels d-f), the demand is elastic ($\beta^{(d)} = -0.4$). The panels on the left (panels a and c) show results for processes in which the demand function does not have any autoregressive features. In the middle (panels b and e), the demand follows AR(1) and ARX(1) processes, and on the right (panels c and f) AR(2) and ARX(2) processes. Demand and supply curves are depicted in solid black and red, respectively. Since the intercept of the supply curve depends on the wind conditions, the dotted red lines represent the 5th and 95th percentiles of hourly wind speed.

## 5. Results

Figure 7 shows the estimated slopes of the demand curve. The true effect is shown by the dashed line. Synthetic demand responds to prices only on the right.

### 5.1. Estimates by identification strategy

**OLS and lagprice IV.** As expected, the OLS regression estimates a positive coefficient if demand is not responsive to prices or if there is no autocorrelation of demand, because it is biased by the effect of demand on price. In the case of autoregressive and price-responsive demand, the OLS regression finds the correct sign, but overestimates the effect. This is because the autoregressive properties of demand accumulate the price response over time. This effect is also evident in the different shapes of the point clouds of the synthetic data (Figure 6). The lag-price IV design allows us to identify the true effect when there is no autocorrelation, but incorrectly identifies a positive slope when there is autocorrelation and no price response. This confirms Angrist and Krueger's (2001) warning against using internal instruments in the presence of serial correlation.

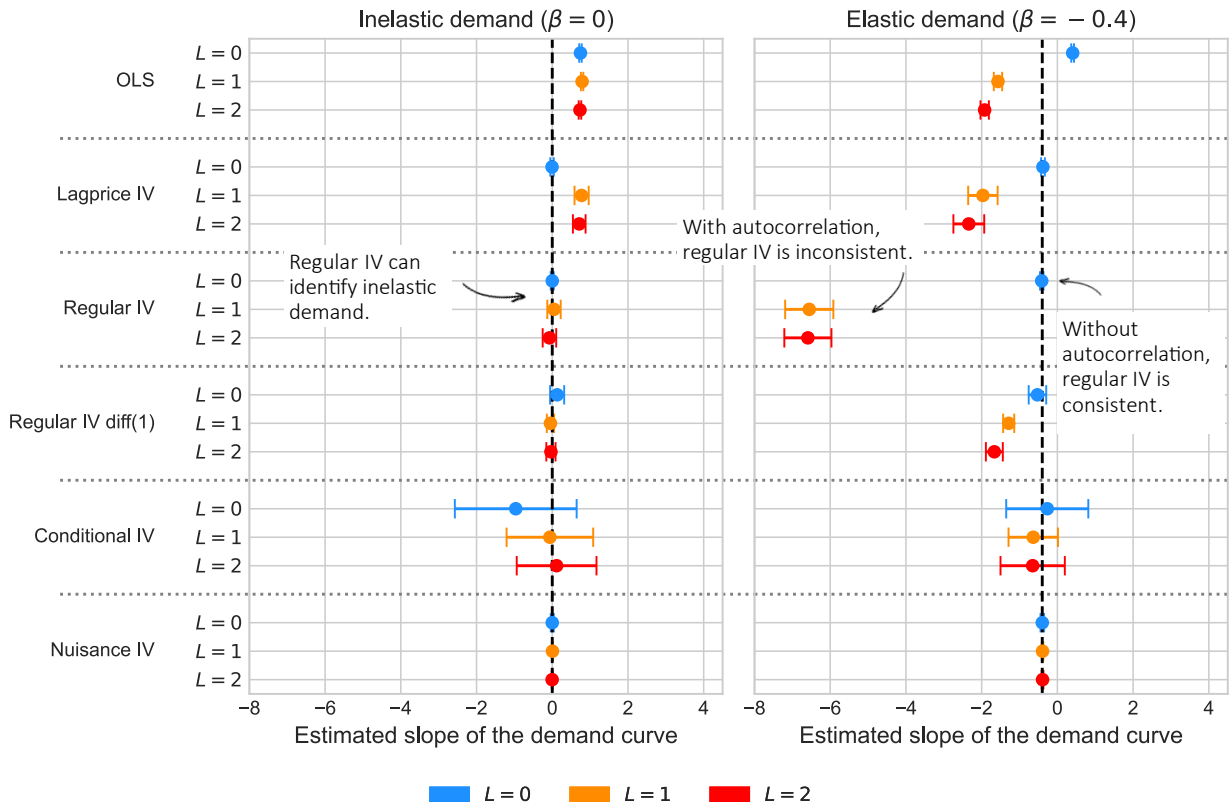

Figure 7. Estimated slope of the demand curve by identification strategy. Demand is inelastic ($\beta^{(d)} = 0$) on the left and elastic on the right ($\beta^{(d)} = -0.4$). The error bars correspond to the 5% significance level obtained from HAC errors. Colors indicate the autoregressive order of synthetic demand.

**Regular IV.** The estimation of the regular IV design is correct when the demand is not responsive to prices or when demand is not autocorrelated. However, once demand exhibits autoregressive behavior and is responsive to prices, the regular IV design leads to an inconsistent estimate. In fact, the regular IV design overestimates the true effect by an order of magnitude. The observations are consistent with the expectations described in Section 0 and show that only the combination of simultaneity and autocorrelation causes the bias. Using first order differences reduces the bias, but not enough to approach the true responsiveness.

**Conditional and nuisance IV.** The two strategies that produce consistent estimates are conditional IV and nuisance IV. In the case of nuisance IV, we always condition on the first two lags ($d_{t-1}$ and $d_{t-2}$), which we know is sufficient to eliminate the bias. In the case of conditional IV, we include the first 26 lags to account for the periodicity of wind. Figure 8 shows the impact of conditioning on an insufficient number of lags: conditional IV fails to identify a meaningful effect. The relatively large number of control variables compared to other estimation methods explains the large confidence interval, making the effect consistent but also not significantly different from zero.

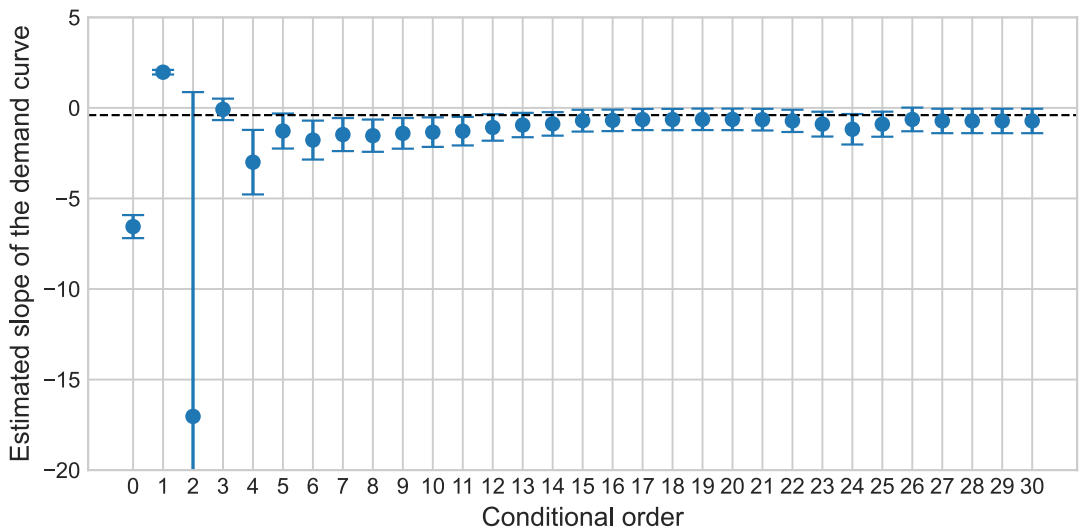

Figure 8: Estimates obtained from conditional IV. The x-axis indicates the number of the lag of wind on which we condition on. The estimation is based on the ARX(1) demand.

## 5.2. Specific cases

**Autoregression of wind.** We argued above that the bias in the estimate should disappear for uncorrelated instruments. To test this, we create a time series in which the wind is shuffled resulting in a wind series without autocorrelation (Shuffled). Figure 9 shows that, as expected, the regular IV estimate using shuffled wind remains consistent even with autocorrelated demand. However, when we use a wind time series with an autocorrelation of order 1 (Synthetic), the regular IV estimate becomes immediately inconsistent. Note that in the case of synthetic demand, the conditional IV finds an effect that is both consistent and significantly different from zero.

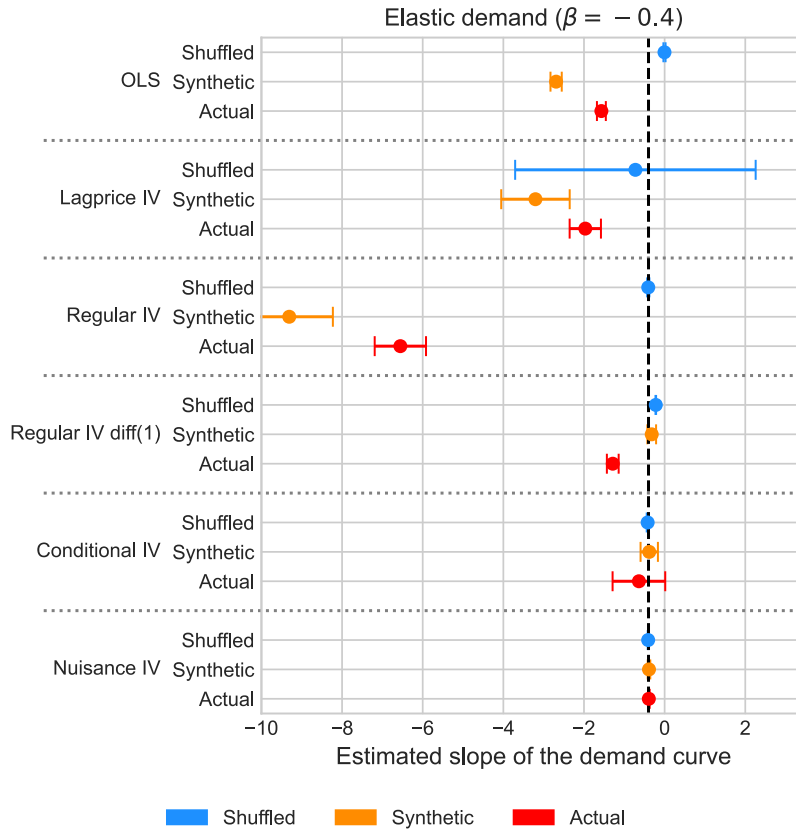

Figure 9: Estimates depending on the autoregressive properties of the instrument. Blue estimates are from shuffled, i.e., uncorrelated wind. Yellow represents synthetic AR(1) wind. Red corresponds to the actual wind. Demand is always elastic ($\beta = -0.4$) and autocorrelated ($L = 1$). Note that the *lagprice* IV does not work for shuffled wind because, by definition, there is no dependency between $p_{t-1}$ and $p_t$ and the price is not a good instrument anymore.

**Autocorrelation of demand.** To better understand the influence of the autocorrelation of demand, we vary the autocorrelation coefficient of the synthetic demand with one lag (Figure 9). Even without autocorrelation (lefthand side) OLS remains inconsistent because of the endogeneity of prices, but all IV approaches provide a consistent estimate of the true demand curve. As the autocorrelation grows (righthand side), IV estimates become increasingly biased, yielding estimates that are (in absolute terms) orders of magnitude larger than the true effect. At an autocorrelation of 0.8 regular IV estimates the effect to be -2, five times the true effect. Throughout, nuisance IV provides consistent estimates.

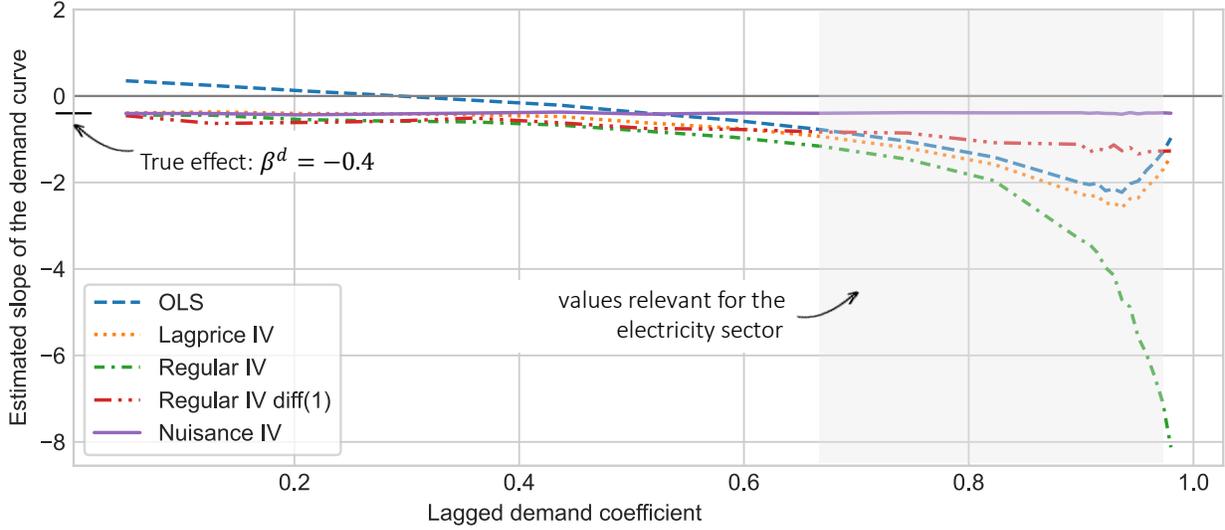

Figure 10. Estimates as a function of the autocorrelation coefficient of demand. The real slope of the demand curve is $\beta^{(d)} = -0.4$. Demand follows an ARX(1) process with a varying autocorrelation coefficient $\alpha_1^{(d)} \in [0,1[$. Only nuisance IV (solid purple) identifies the real effect independently of $\alpha_1^{(d)}$.

## 5.3. Magnitude of the Thams bias

**Thams bias.** Thams et al. (2022) show that the bias induced by the autocorrelation depends on the autocorrelation coefficients of the instrument and the demand only, if both time series follow an autoregressive process of order 1 (see section 3.2). First, we examine the extent to which the Thams bias follows the same functional relationship in our case of simultaneous price and demand determination. To do this, we use the synthetic wind time series for which the autoregressive properties are known. As can be seen in panel (a) of Figure 11, the prediction based on the formula for the Thams bias is in line with the measured estimate (Table 1). However, this is a theoretical exercise since these true autoregressive coefficients are unknown. Calculating the bias with estimated autoregressive coefficients of the demand time series (Table 2) is not a good predictor for the actual bias. The observed bias (blue solid line) in the case of an autoregressive process of order 2 for the demand and the actual observed wind series clearly follow the same functional form. A prediction based on the following formula comes very close to the actual observations: $\hat{\beta} = \frac{1}{1-\sum_{i=1}^{L} \alpha_i^{(d)} \sum_{i=1}^{M} \alpha_i^{(w)}} \beta$, where $\alpha_i^{(d)}$ and $\alpha_i^{(w)}$ represent the autocorrelation of lag $i$, $L$ the ARX degree of demand (dependent variable) and $M$ the AR degree of wind (instrument).

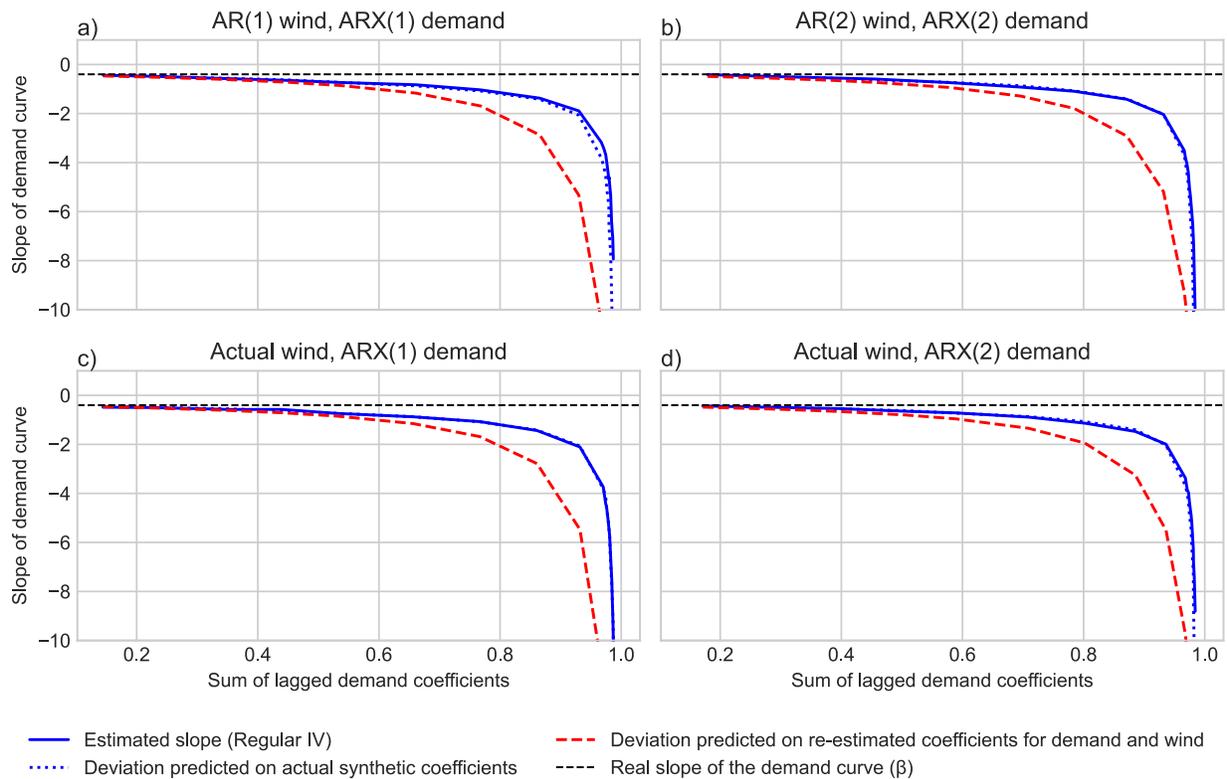

Figure 11: Predictions of the Thams bias compared to the actual estimates. In the upper row we use a synthetic wind series for which we know the exact autoregressive properties, the lower row the actual wind as used for the main results of this paper. Panels on the left model electricity demand as an ARX(1) process, on the right as an ARX(2) process.

## 6. Discussion

**Consequence**. As a consequence of our results, we should be warned against using a regular IV design to estimate the short-term elasticity of electricity demand. If a policy maker, system operator or researcher were to use the inflated estimate further, she could severely overestimate the ability of flexible demand to support the integration of renewables and reduce market power and, in the worst case, even plan for insufficient back-up capacities. If at all, a regular IV design may only be justified for an energy economist who wants to show that electricity demand is unresponsive to price, as Fabra et al. (2021) aim for.

**Cumulative response.** If not the short-term elasticity of demand, which response – in energy economic terms – does a regular IV measure? We argue that the estimate must not be understood as pure inconsistency. The wind does not change instantaneously. As a result, wind-driven supply-side effects persist for extended periods, and prices would tend to be low or high over extended periods as well. The autoregressive behavior of demand accumulates the effect over these periods (see Figure 12). This means that even a small response to prices can lead to a significant reduction in electricity demand over time – which is picked up by regular IV regression.

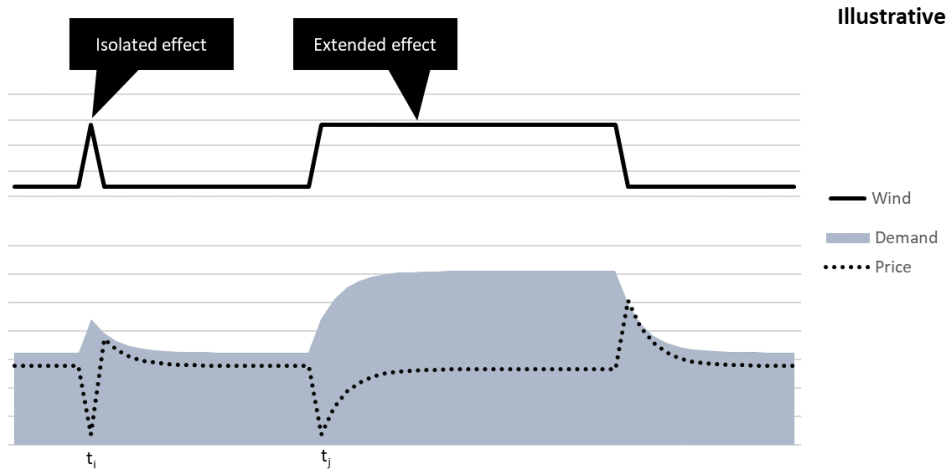

Figure 12: Illustrative effect of an isolated, i.e., non-correlated wind peak, compared to extended, i.e. autocorrelated, high wind period. The non-correlated wind peak only increases demand around $t_i$ while the demand response builds up over a longer time if the effect is extended. The estimate at $t_j$ needs to be biased.

# 7. Conclusions

**Summary.** This paper showed the difficulties of using instrumental variable (IV) regressions to estimate the response of electricity demand to high-frequency price signals. We generate synthetic data where the demand curve follows an autoregressive process and prices are endogenously determined by calculating the market equilibrium. We show that regular IV estimation generally leads to inconsistent estimates of the slope of the demand curve. We show that the degree of inconsistency, the Thams bias, could be theoretically predicted if the autoregressive properties of the demand curve were known. In empirical applications, however, this is not the case. Therefore, an ex-post correction is not possible. Finally, we argue that a regular IV estimate should be interpreted as the cumulative effect of the price response, but not as the very short-term responsiveness. Our work also discussed two strategies that provide consistent estimates of the short-term elasticity of electricity demand: nuisance and conditional IV. The former includes lags of the dependent variable as controls, while the latter includes lags of the instrument. As the strategies are easy to implement in empirical applications, we recommend that researchers use both strategies.

**Further research**. We see value in further research in three areas: synthetic data generation, theory, and applications. Synthetic data generation can be extended to include non-linear functional relationships and autocorrelation of supply, more sophisticated forms of demand response, and high-frequency cross-sectional data. Theoretical work can investigate if and how conditional and nuisance IV can be adjusted to capture heterogeneous effects and provide a proven closed form for the magnitude of the Thams bias for AR(p) processes. Issues that need to be investigated further to transfer the method and results to a real application in the electricity sector include the correct specification of conditional IV in the presence of real demand, the behavior in case of shorter time series (small T) and the extent to which wind time series could be aggregated to reduce autocorrelation enough to justify the use of a regular IV design.

## Funding

This work was supported by the German Federal Ministry of Education and Research through the ARIADNE Project (FKZ 03SFK5K0).

## Acknowledgement

We thank Oliver Ruhnau, Tarun Khanna, William Lowe, Nikolaj Thams, and Jonas Peters for valuable feedback and insightful discussions. We thank Jorge Sánchez Canales for superb research assistance.